# A unique metal-semiconductor interface and resultant electron transfer phenomenon

S. L. Taft*

*Flowing Water Consulting, PO Box 709 Geneva, OH 44041-0709*

An unusual electron transfer phenomenon has been identified from an n-type semiconductor to Schottky metal particles, the result of a unique metal semiconductor interface that results when the metal particles are *grown from* the semiconductor substrate. The unique interface acts as a one-way (rectifying) open gateway and was first identified in reduced rutile polycrystalline titanium dioxide (an n-type semiconductor) to Group VIII (noble) metal particles. The interface significantly affects the Schottky barrier height resulting in electron transfer to the metal particles from the reduced rutile titanium dioxide ($TiO_2$) based on their respective work functions. The result is a negative charge on the metal particles which is of sufficient magnitude and duration to provide cathodic protection of the metal particles from surface oxidation.

## I. INTRODUCTION

The data supporting an unusual electron transfer phenomenon is significant for its highly unusual negative binding energy (B.E.) shifts below zero valence energies observed for rhodium particles on reduced rutile titanium dioxide ($TiO_2$). The discovery, made more than thirty years ago, was the result of the initial failure to make a rhodium-doped $TiO_2$ material ($Ti_{0.99}Rh_{0.01}O_2$) using the Pechini[1] process to reduce the band gap for a $TiO_2$-based photo-electrochemical material. After formation of the resin intermediate from Pechini process, the material was subjected to high temperature oxidation and reduction in excess of 1400°C. X-ray photoelectron spectroscopy (XPS) was chosen to help verify Rh substitution into the $TiO_2$ lattice by identifying its oxidation state. The B.E. for Rh $3d_{5/2}$ for materials prepared with three different oxidation and reduction temperatures (1500° C, 1600° C and 1650° C) were consistently far below the reference of 307.1 electron volts (eV) for $Rh^0$ (301.2 eV, 302.2 eV, and 302.6 eV, respectively). This corresponds to negative B.E. shifts of 6.1, 5.1, and 4.7 eV. However, the B.E. for Ti $2p_{3/2}$ did not exhibit the same negative shifts and remained within expected values for a $Ti^{4+}/Ti^{3+}$ mixed valence oxide. This data was highly unusual and prompted additional study of the materials. Subsequent examination by scanning electron microscopy (SEM) using energy dispersive x-ray (EDX) analysis found that rhodium was observed in discrete particles on the surface of the polycrystalline $TiO_2$ material, particles that were *grown from* the $TiO_2$. The particles were both regularly and irregularly-shaped microspheres from approximately 1 to 10 micrometers (μm) in size. Two additional materials (1.7 mole % rhodium-doped $TiO_2$) with reduction temperatures of 1400° C and 1500° C had smaller particles of rhodium, approximately 1 μm and less, and smaller negative B.E. shifts of 0.5 and 1.2 eV, respectively.

A literature search at the time (1981-1982) found 3 instances where negative B.E. shifts of metals were reported: 1) surface atom core-level shifts (SCLS), both positive and negative, relative to bulk[2]; 2) metal alloys[3]; and 3) sub-monolayer thin film coverage and small clusters [4,5,6]. In the case of SCLS, Johansson and Martensson[2] attributed the shifts to the difference in potential experienced by surface atoms relative to bulk due to the lower coordination number. Their calculated shifts indicated negative shifts on the order of 0.3 eV for iridium and platinum. Kleiman et al [3] studied B.E. shifts for various ratios of Pt-Cu alloys and found that while Cu $2p$ shifted up to -0.7 eV, no shift was observed for Pt 4*d* lines. In regard to sub-monolayer thin film coverage and small clusters, both positive and negative B.E. shifts were observed relative to bulk materials. Both Mason and Baetzold[4] and Takasu et al[5] measured positive shifts for small clusters, 2.5 eV for silver on carbon, and 1.6 eV for palladium on amorphous silver, respectively. Oberli et al[6] measured negative B.E. shifts of 0.7 eV for the smallest gold clusters on carbon. Due to the significant differences in material preparation and material-particle morphology, it was concluded that none of these instances provided an appropriate explanation for the negative B.E. shifts observed in this study. Additionally, the magnitude of the largest negative B.E. shifts observed in this study exceeded other reported shifts, often by more than a full order of magnitude.

Negative B.E. shifts were also reported for Group VIII metals on reduced $TiO_2$ catalyst supports during investigations of unusual catalytic activity attributed to Strong Metal Support Interactions (SMSI). While some of the XPS studies of SMSI materials in the early 1980's reported negative B.E. shifts of the supported metal particles, in many instances, the negative shifts were more consistent with those observed for sub-monolayer coverage and small clusters. It should also be noted that since high temperatures are contra-indicated for maintaining optimal surface area for catalysis, the bulk of the high temperature reduction temperatures during SMSI investigations rarely exceeded 500° C. However, while there did not appear to be a direct correlation with SMSI investigations, the similarity of materials was intriguing and helped to steer a series of experiments to further elucidate the cause of the negative B.E. shifts.

This paper reviews the original experiments designed to elucidate the cause of the unusual negative B.E. shifts. The paper will also re-evaluate data from the original experiments with more recent information obtained from the literature to support the unusual electron transfer phenomenon, referred to in this paper as the open gate phenomenon.

## II. EXPERIMENTAL

The materials (Rh, Pt, and Ir in $TiO_2$) were prepared using two different processes prior to high temperature oxidation and reduction. The initial series of materials were prepared using the Pechini process which is based on U.S. Patent 3,330,697 following Example I of the patent. The resin intermediate was prepared by adding the desired amount of Group VIII metal from 1% to 2% mole percent ($Ti_{100-x}M_xO_2$) followed by calcination (oxidation) air in temperatures ranging from 400-500° C. The subsequent material was ground with a mortar and pestle to obtain a powder which was then subjected to high temperature oxidation (100% $O_2$) and reduction (100 $H_2$, 1% and 5% $H_2$ in Ar) in a tube furnace.

A metal salt impregnation-evaporation process, following the procedure for catalyst preparation outlined by Tauster et al[7] (first identified SMSI), was also used as a substitute for the Pechini process step. The materials were prepared by impregnating $TiO_2$ with metal salt solutions (0.5 weight percent), followed by air drying prior to high temperature oxidation and reduction. In all instances, the final materials were stored in plastic or glass vials under normal atmospheric conditions with no special handling precautions to maintain an inert environment prior to XPS or SEM/EDX analysis.

SEM analysis was performed on a JEOL JSM-35C scanning electron microscope equipped with a solid-state backscatter electron detector and Princeton Gamma-Tech x-ray detector. The samples were mounted using a conductive carbon paint/paste. A conductive carbon coating (via carbon evaporation) was not applied prior to analysis since the materials were generally sufficiently conductive to eliminate charging. Microprobe x-ray analysis (SEM/EDX) and backscattered electron imaging using Z-contrast (imaging based on percentage of backscattered electron relative to atomic number) provided the ability to identify the location of Group VIII metals relative to $TiO_2$.

Two different XPS spectrometers were used for analysis, a PHI (Physical Electronics, Eden Prairie, MN) spectrometer with Auger and XPS capabilities and a Varian (Varian Analytical

Instrument Devices, Palo Alto, CA) spectrometer. XPS data was typically collected within a few days of material preparation. Charge correction was based on adventitious carbon (C1s) at 284.6 eV. Except where noted, ion (Ar) sputtering was not employed prior to analysis. However, reference data was measured post ion sputtering on pure foil and powders obtained from Alpha Products.

## III. RESULTS

*First Incidence*

The unusual XPS data was initially observed for Rh$3d_{5/2}$ in a material prepared with high temperature (1500° C) oxidation (100% $O_2$) and reduction (100% $H_2$) post-preparation of a Pechini resin intermediate to obtain a 1 mole percent rhodium substitution for $TiO_2$ ($Ti_{0.99}Rh_{0.01}O_2$). Upon observing the initial raw data, the spectrometer was recalibrated and the sample was ion sputtered for five minutes before the second set of data was obtained and which again, confirmed the highly unusual negative B.E. shift below reference data for $Rh^0$ of 307.2 eV.[8] The data in Table 1 is the raw XPS data with no charge correction applied.

**Table I. "First Incidence" Initial Data**

| 1400° C (oxidation & reduction) | (Pre-sputter) raw data | (Post-sputter) raw data |
|---|---|---|
| C$1s$ | 286.7 eV | 284.9 eV |
| Rh$3d_{5/2}$ | 303.3 eV | 303.0 eV |
| Ti$2p_{3/2}$ | 458.7 eV | NM |
| Ti$2p_{1/2}$ | 464.6 eV | NM |

NM- not measured

Ruling out instrument error, the issue of charge correction presented two possible options. The first used the standard adventitious C$1s$ for correction (284.6 eV) which resulted B.E. data for Ti$2p_{3/2}$ of 456.6 eV. Based on the high temperature reduction, this was fairly consistent with the presence of $Ti^{3+}$. The second option of basing charge correction on Ti$2p_{3/2}$ also did not resolve the substantial negative B.E. shift for Rh$3d_{5/2}$. However, regardless of which charge correction was used, the magnitude of the negative shifts for Rh$3d_{5/2}$ far exceeded either charge correction option. It was therefore concluded that using the standard adventitious carbon charge correction protocol provided the only possible option (Figure 1).

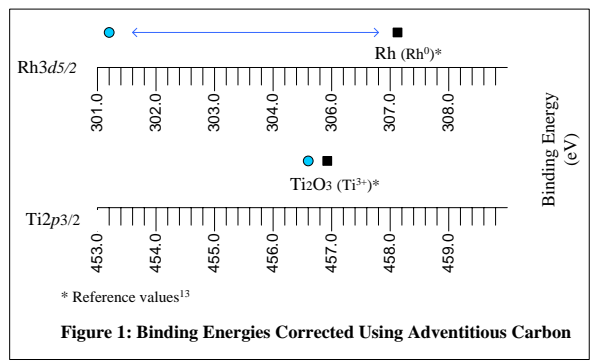

**Figure 1: Binding Energies Corrected Using Adventitious Carbon**

Two additional materials (1600° C and 1650° C oxidation and reduction) were analyzed which again confirmed substantial negative B.E. shifts for rhodium. Again, using adventitious carbon for charge correction, the data was consistent for the presence of $Ti^{4+}$ and $Ti^{3+}$, from mixed titanium oxides (Figure 2).

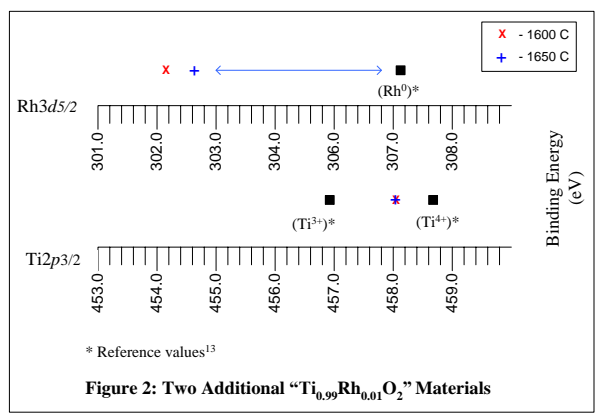

**Figure 2: Two Additional "$Ti_{0.99}Rh_{0.01}O_2$" Materials**

Examination of the materials using SEM with EDX and backscattered imaging did not indicate that rhodium was within the rutile structure. Instead, it was observed in discrete regularly and irregularly shaped microspheres, often at the grain boundaries of $TiO_2$. The particles were approximately 1 to 10 μm in size. The irregular shape of some of the particles was often due to the presence of a flat surface facet interposed onto the more-or-less spherical particles. It was concluded that substitution of rhodium into the TiO$_2$ lattice producing a rhodium-doped material was not achieved.

*Material Process Variations for Pechini-based Materials*

The next series of experiments explored two aspects of material preparation in an attempt to understand how the metal particles which exhibited the unusual negative B.E. shifts were formed: 1) substitution concentration, and 2) reduction temperature and environment (Table 2). High temperature oxidation remained a constant (1650° C in O$_2$). However, prior to any XPS analysis of the materials, the B.E. for Rh$3d_{5/2}$ of a rhodium metal reference (Rh$^0$) was determined. This reference value (307.1 eV) was used to determine B.E. shifts for all subsequent experiments for rhodium.

**Table 2. Additional Pechini-based Materials**

| Sample ID | Substitution (mole %) | Reduction Preparation* |
|---|---|---|
| P1A | 1.0 | No reduction step |
| P1B | 1.0 | 600° C in H$_2$ |
| P2A | 2.0 | No reduction step |
| P2B | 2.0 | 600° C in H$_2$ |
| P3A | 1.7 | No reduction step |
| P3B | 1.7 | 1500° C in Ar |
| P3C | 1.7 | 1400° C in Ar |

*Ar is a milder reducing environment than H$_2$

Both positive and negative B.E. shifts for Rh$3d_{5/2}$ were observed in the materials (Figure 3). As expected, XPS examination for Rh$3d_{5/2}$ B.E. confirmed the presence of Rh$_2$O$_3$ (Rh$^{3+}$) in the materials which were oxidized only (P1A, P2A, and P3A) with positive B.E. shifts (+2.0 and +2.2 eV) relative to Rh$^0$ at 307.1 eV. The data for the materials reduced in H$_2$ at 600° C (P1B and P2B) exhibited smaller positive B.E. shifts (+0.3 and +0.4 eV respectively) indicating reduced rhodium metal with surface oxidation. The materials reduced in Ar at significantly higher temperatures (1500° C for P3B, 1400° C for P3C) exhibited smaller negative B.E. shifts (-1.2 eV and -0.5 eV, respectively) than observed in the initial materials.

Examination by SEM/EDX of fully oxidized material (P3A) showed large agglomerated particles up to 20 μm containing rhodium mixed in with the TiO$_2$. The particles in the materials exhibiting negative B.E. shifts (P3B and P3C) were significantly smaller in the range up to 1 μm but were also regularly and irregularly shaped microspheres, consistent with the initial materials.

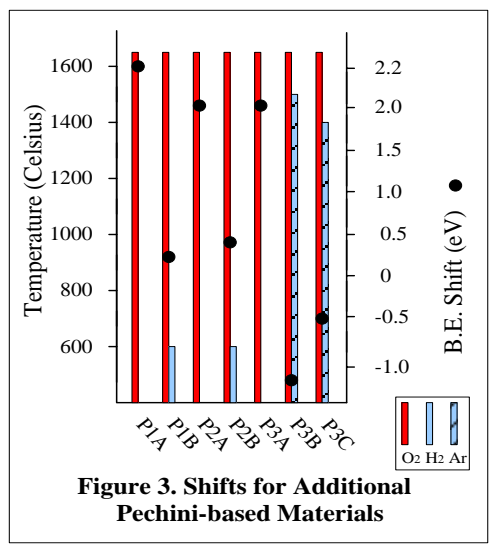

**Figure 3. Shifts for Additional Pechini-based Materials**

*Material Process Variations Based on Metal Salt Impregnation-Evaporation Process*

While extreme oxidation and reduction temperatures (greater than 1400° C) produced negative B.E. shifts in Pechini-based materials, the high temperatures also produce large particle sizes and size ranges. In an attempt to determine whether negative B.E. shifts were specifically linked to the Pechini-process precursor and also minimize particle size and range, materials (0.5% Rh/TiO$_2$) were prepared using a metal salt impregnation-evaporation process. The materials were subsequently subjected to a series of oxidation and reduction temperatures and environments to determine the approximate lower temperature thresholds and reducing atmospheres required to obtain materials that exhibit negative B.E. shifts (Table 3).

Figure 4 illustrates a link between the lowest oxidation temperature threshold and reduction temperature and atmosphere relative to obtaining negative B.E. energy shifts. The shifts were calculated relative to the rhodium foil reference (Rh$^0$ $3d_{5/2}$ at 307.1 eV). An approximate lower temperature limit for oxidation and reduction is 800° C although a more severe reducing atmosphere (5% H$_2$) can lower the oxidation threshold (MS-9 and MS-10) provided the minimum reduction temperature is approximately 800° C. Additionally, this series of experiments confirms that the oxidation and reduction conditions are critical to obtaining the negative B.E. shifts and not specifically dependent upon the Pechini-based precursor material.

**Table 3. Oxidation and Reduction Temp. and Reduction Environment Relative to B.E. Shift**

| Sample ID | Oxidation Temp. (°C) | Reduction Temp. (°C) | (% H$_2$ in Ar) | B.E. Shift (eV)[a] |
|---|---|---|---|---|
| MS1 | 450 | --- | --- | ND[b] |
| MS2 | 800 | --- | --- | +1.6 |
| MS3 | 450 | 850 | 1 | ND[b] |
| MS4 | 600 | 800 | 1 | +1.2 |
| MS5 | 800 | 800 | 1 | -0.7 |
| MS6 | 800 | 1150 | 1 | -0.7 |
| MS7 | 1020 | 850 | 1 | -0/7 |
| MS8 | 450 | 525 | 5 | ND[b] |
| MS9 | 500 | 800 | 5 | -0.8 |
| MS10 | 500 | 1000 | 5 | -1.6 |
| MS11 | 1000 | 1000 | 5 | -0.7 |

[a] Metal reference: Rh$^0$ $3d_{5/2}$ at 307.1 eV
[b] Not Detected by XPS although presence confirmed by XRF.

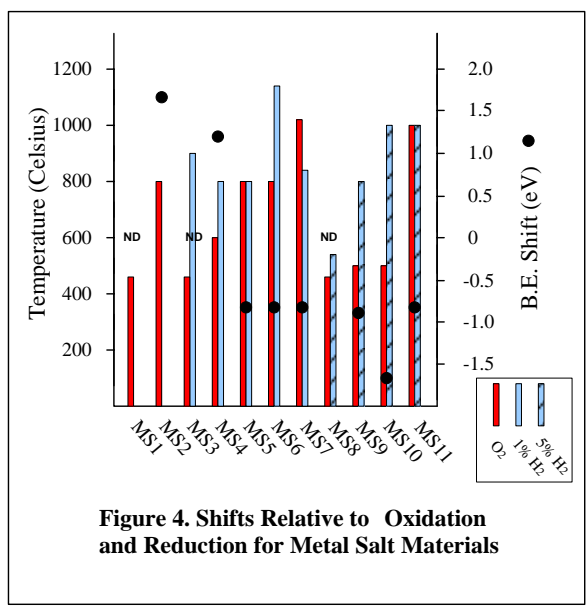

**Figure 4. Shifts Relative to Oxidation and Reduction for Metal Salt Materials**

It should be noted that while no rhodium was detected by XPS for MS1, MS3, and MS8, x-ray fluorescence analysis confirmed the presence of rhodium in the samples. This indicated that the metal is not present in sufficient quantity on the surface of the TiO$_2$ crystals for detection by XPS.

SEM/EDX analysis and SEM-backscatter imaging (Figure 5) confirmed randomly distributed rhodium microspheres, 0.2 μm and less, which supports the assumption that lower temperatures produce smaller particle sizes. The smaller particle sizes may also contribute to the smaller magnitude B.E. shifts compared to the larger particle sizes observed with the Pechini-based materials.

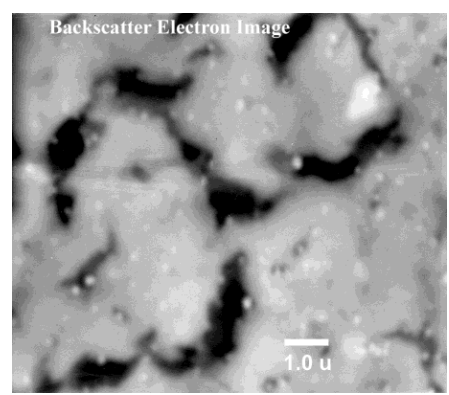

**Figure 5. Rhodium Particle Location Using Z-Contrast of Backscatter Electron Image of MS10**.

Substitutions of two other Group VIII metals (Ir and Pt) were also prepared using oxidation and reduction temperatures (800° C oxidation, 1000° C reduction in 1% H$_2$) which were successful for 0.5% Rh/TiO$_2$. Negative B.E. shifts were observed for both Ir and Pt (Table 4) and were within the range consistent with respect to particle size for 0.5% Rh/TiO$_2$.

**Table 4. B.E. Shift Data for Pt and Ir Substitutions**

| Sample ID | Material | B.E. Shift (eV)* |
|---|---|---|
| MS12 | 0.5% Pt/TiO$_2$ | -0.3 |
| MS13 | 0.5% Pt/TiO$_2$ | -0.7 |
| MS14 | 0.5% Ir/TiO$_2$ | -0.7 |

*Referenced to Pt$^0$ $4f_{7/2}$ at 71.1 eV or Ir$^0$ $4f_{7/2}$ at 60.9 eV

*Persistence of Negative B.E. Shifts*

Persistent negative shifts without the formation of surface oxides were observed many weeks after material preparation. This was considered highly unusual without maintaining an inert storage atmosphere for the materials. Five different materials (four Rh/TiO$_2$ and one Ir/TiO$_2$) had persistent negative B.E. shifts up to 169 days post preparation. Figure 6 illustrates the longevity of B.E. shift of the materials analyzed for this study. Of particular interest is that surface oxidation does not occur where negative B.E. shifts are observed.

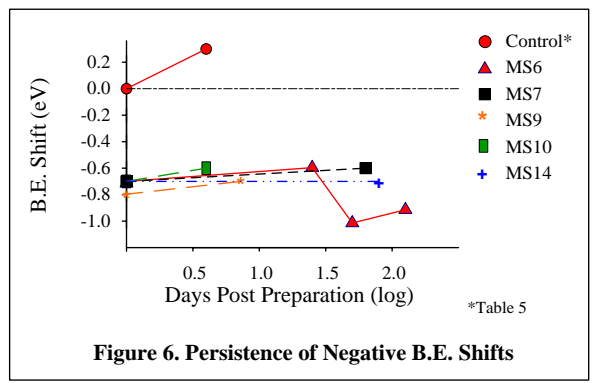

**Figure 6. Persistence of Negative B.E. Shifts**

The last XPS measurement for MS6 was 169 days post preparation. No rhodium was detected; however, neither was titanium. The *C 1s* peak indicated an organic contamination on the surface of the material.

*Rhodium Powder Control Experiment*

A series of samples were prepared and analyzed by XPS as a rhodium powder control experiment (Table 5) in order to illustrate the effects of high temperature oxidation, reduction, and air exposure at ambient atmospheric conditions on pure rhodium powder.

**Table 5. B.E. Shift Data for Rhodium Powder Control Experiment**

| Sample ID | Preparation | Analysis Details | B.E. Shift (eV) |
|---|---|---|---|
| A | As received | Not sputtered | +0.4 |
| B | 900° C in air | Day 1 | +1.4 |
| C | 900° C/air, 800° C/% H$_2$ | Day 1 | 0.0 |
| C | | Day 5 | +0.4 |
| C | | Day 5, post sputter | 0.0 |

The experiment confirmed: 1) the presence of surface oxides on pure rhodium metal powder without maintaining an inert atmosphere for storage; 2) high temperature exposure in air oxidized the surface of rhodium metal powder; 3) high temperature exposure to a reducing atmosphere removed surface oxides as did ion sputtering prior to XPS analysis, and; 4) surface oxides readily formed on metal powder within days of exposure to atmospheric air. The experiment also illustrated and confirmed that the high temperature oxidation and reduction required to produce negative B.E. shifts in Rh/TiO$_2$ did not produce the same shifts in rhodium powder alone.

## IV. DISCUSSION

Many questions remained unanswered when the original research ended in 1982. With considerably more information available regarding TiO$_2$ owing to its applications in catalysis, photo-catalysis and sensors, the original data was revisited coincident with a review of the literature since that time. With one exception, the review still does not report any instances of similar experimentation and data. However, the review does provide information which contributes to an understanding of the original data.

One area of interest specifically relates to the lower threshold temperature to produce the negative B.E. shifts. In 1982, the threshold temperature of approximately 800° C was thought to be lower than the presumed threshold of approximately 1000° C for the anatase-rutile transformation for TiO$_2$. A more recent paper by Gouma et al[9] studying anatase-rutile transformation referenced research dating back fifty years to a paper by Shannon and Pask[10] with data that indicated that the transformation can occur as low as 400° C depending upon the synthesis method, the atmosphere, and the presence of other ions. They noted that reduction increases the rate and lowers the transformation temperature due to oxygen vacancies. This confirmed a study by Iida and Ozaki[11] where rutile transformation was enhanced by a reducing environment (argon, hydrogen, and a vacuum). Their studies also found that additions of transition metals can promote the transformation at lower temperatures. Therefore although the specific causal link is not currently understood, the threshold temperature required to obtain negative shifts is linked to the transformation to the rutile structure of $TiO_2$.

Although there did not appear to be a direct correlation with SMSI investigations, a review of SMSI-related literature also provides intriguing information that supports the original data. SMSI-related research grew significantly in the mid-1980's beyond the original catalysis investigations. This included experiments designed to test for electron transfer of metals deposited onto single crystal $TiO_2$ using surface analysis techniques as well as investigations into other reducible oxides and the use of dopants to explore the semiconductor properties of reduced $TiO_2$. Unfortunately, no single electronic-based hypothesis could explain the often mixed and confusing results with the widely ranging materials preparations from polycrystalline catalysts to single crystal in-situ preparations. An alternative explanation evolved based on data that indicated encapsulation of the metal particles by a TiO suboxide layer which is the current preferred explanation for SMSI. However, a few more recent catalysis studies, including gold on reduced $TiO_2$, provide support for an electronic basis for SMSI. It is some of these studies which also provide intriguing information that lends itself to the original data reviewed in this paper.

In a 2001 study of $Au/TiO_2$ catalysts (prepared using a metal salt impregnation-evaporation process, not single crystal in-situ preparations), Akita et al[12] found preferential nucleation of gold clusters on rutile structures prepared at different calcination temperatures. Their data indicated preferential nucleation of Au particles at grain boundary interfaces and that it initiates during calcination (high temperature oxidation). They also noted a "significant difference between catalyst prepared by deposition precipitation [polycrystalline] and a model catalyst made by evaporation of gold onto the surface of a single crystal of rutile."

In 2003, Eider and Kramer[13] investigated the structural and electronic effects of high reduction temperatures on $Pt/TiO_2$ catalysts. The reduction temperatures for this study ranged up to 800° C, higher than many previous studies and which is a temperature sufficient for complete transformation to rutile. Their findings were significant in regard to the effects of temperature on grain size, surface area, and conductivity. They concluded that the "higher density of charge carriers causes a narrowing of the space charge region at the metal support phase boundary" and that the "charge transfer necessary to align the Fermi levels of the two phases becomes larger which most likely affects the catalytic properties of these catalysts."

In a 2005 paper, Iddir et al[14], also studying polycrystalline $Pt/TiO_2$ catalysts observed an "unexpected" preferential nucleation for Pt particles on rutile over anatase with reduction and that the distribution of Pt seemed independent of surface orientation of $TiO_2$ particle with a higher concentration at the $TiO_2$ particle edges. They also observed a "nearly perfect spherical shape and thus has a point-like contact with titania [$TiO_2$]" which as they explained "shows the complex inter-dependence between the Pt particle shape, Pt-$TiO_2$ interface extent, $TiO_2$ phase, surface crystallographic orientation, and possibly the local density of oxygen vacancies." Although their investigation did not include any XPS analysis, their results support the effects of charge transfer on the morphology of the supported metal particles.

In regard to the referenced SMSI papers, each provides interesting data in support of the original data presented here as well as reinforcing the complexity of polycrystalline $TiO_2$ in regard to electronic interactions. However, with one exception, none of the SMSI-related literature reports any instances of similar experimentation and data. The single exception is a paper published in 1986 by Spichiger-Ulmann et al.[15] Their materials were prepared as $Pt/TiO_2$ catalysts for XPS analysis using salt impregnation-evaporation of thick films of $TiO_2$ in the range of 10 to 15 μm followed by air calcination and reduction in Ar. Subsequent XPS analysis reported negative binding energy shifts from 0.2 to 0.6 eV for Pt particles (less than 1 μm) below their $Pt^0$ reference prepared as a thick film on titanium. Additionally, the negative B.E. shifts were "persistent even after exposure of the samples to air for several days." Negative B.E. shifts were observed for up to 75 days for one sample and for another, showed a positive shift back to a B.E. relative to the standard (Pt $4f_{7/2}$ at 71.1 eV) at four months (Pt1 and Pt2, respectively in Figure 7).

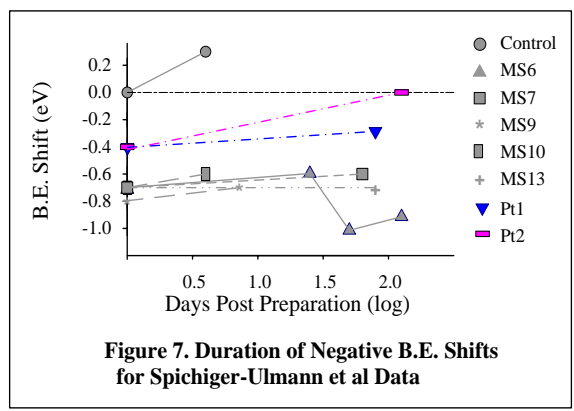

**Figure 7. Duration of Negative B.E. Shifts for Spichiger-Ulmann et al Data**

It is noted that the materials were prepared as platinized (via an aqueous impregnation) $TiO_2$ thick films resulting in polycrystalline $TiO_2$. Additionally while the oxidation and reduction temperatures (450° C and 550° C, respectively) were below the lower thresholds identified in this study, rutile transformation (although not specifically determined) is nevertheless possible based on Shannon and Pask[12] and Iida and Ozaki[13]. It is also noteworthy that the magnitude of negative B.E. shifts is consistent with the data presented in this paper given the size of the Pt particles.

Based on all of this information, a clearer understanding has evolved for the conditions required to prepare materials which exhibit negative B.E. shifts: 1) material preparation that contributes to intimate mixing of metal oxides with $TiO_2$ and subsequent mobility that accompanies high temperature oxidation; 2) sufficient reduction of $TiO_2$ to an n-type semiconductor, and 3) transformation of $TiO_2$ to the rutile structure. The high temperature oxidation allows the metal to grow from the $TiO_2$ substrate (regardless of the initial pre-oxidation preparation) as opposed to a deposition onto a $TiO_2$ support. This implies an inherent difference between a metal-substrate interface as opposed to a metal-support interface and is critical in understanding the underlying mechanism of the negative shifts.

Group VIII metals on reduced $TiO_2$ are Schottky metals on an n-type semiconductor. In metal-semiconductor contact theory, thermodynamic equilibrium assumes that the electrochemical potential is uniform throughout the system. A potential forms at the junction of a metal and semiconductor, a Schottky barrier, based on their respective differences in electrochemical potentials. The ideal barrier height ($\phi_B$) is determined by:

$$\phi_B = \phi_M - \chi_S$$

Where:    $\phi_M$ is the work function of the metal
$\chi_S$ is the electron affinity of the semiconductor

As contact is made, electrons will flow initially to the material with the smaller potential. For $TiO_2$ which is a strongly ionic semiconductor, the barrier height will be equal or proportional to the difference in work functions. For Rh, Ir, and Pt in contact with $TiO_2$, the contact potential is the difference in work functions: 4.6 eV[16] for reduced $TiO_2$, 4.98 eV for Rh, 5.61 eV for Ir (average for 100, 110, and 111 crystal planes) and 5.64 eV for Pt.[17] The result is electrons transferred to the metal particles. However, as the barrier forms, the flow of electron slows and eventually stops due to the formation of a depletion (of charge) zone, the charge neutral layer (CNL). In traditional semiconductor fabrication and applications, the presence of a thin layer of interfacial oxides at the metallization layer-semiconductor interface is a significant factor determining the actual barrier height.

However, the interface where the metal is grown from the semiconductor itself is not as straightforward. Additionally, the semiconductor in this situation is also not a single crystal. The semiconductor is a polycrystalline material with grain boundaries and defect sites which affect electrical properties relative to assumptions made for single crystals. While no metal was detected within the rutile lattice, this was also not confirmed using another more precise analytical analysis and therefore some level of metal doping of the rutile lattice may also contribute to the mechanism. All of this adds to the complexity of understanding electron transport and conductivity of reduced $TiO_2$.

Nevertheless, based on metal-semiconductor contact theory, the junction is a rectifying junction. The electrons flow one-way into the (Schottky) metal particles. This is key to understanding the cause of the negative B.E. shifts and is best understood by recognizing what affects binding energies as well as how binding energies are measured in an XPS spectrometer.

The binding energy of an electron depends on the energy level from which it originates ($1s$, $2s$, $2p_{3/2}$, etc…) but also on the oxidation state of the atom as well as the local chemical and physical environment. Binding energy ($E_{binding}$) is derived from the measurement of the kinetic energy ($E_{kinetic}$) of the electron in the spectrometer are based on the following:

$$E_{binding} = E_{photon} - E_{kinetic} - \phi$$

Where:    $E_{photon}$ is the energy of the excitation radiation
           $\phi$ is the work function of the spectrometer (not the material/sample analyzed)

The key assumption is that both the sample and the spectrometer are at the same potential (grounded to one another). Herein is the explanation for the negative B.E. shifts. The metal particles are in essence isolated from the $TiO_2$ substrate and therefore from the spectrometer due the rectifying interface between the reduced $TiO_2$ and the metal particles. The negative B.E. shifts are therefore the result of the local environment alone created by the transfer of electrons from reduced $TiO_2$ (n-type semiconductor) and the metal particles (Schottky metal). The negative charge on the particles *is* the local environment and explains why corresponding shifts were not observed for titanium binding energies.

The negative charge is established during high temperature reduction and is of sufficient magnitude to affect morphological changes in the particles exhibiting the negative charge. Due to Gauss' law, the preferred shape is a sphere for a point charge to equalize charge interaction. It is proposed that the metal microspheres are therefore conducting spheres where the charge resides on the surface, instead of only at the interface. However, at a given magnitude of charge on a microsphere relative to its size, the energy balance tips in favor of the crystal lattice structure. This explains the observations of crystal facets and edges protruding from some of the very large microspheres. This also implies a limiting factor as to the size of the charge on the particles-microspheres which in itself, is an apparent contradiction to what is traditionally assumed regarding the amount of charge transferred for $TiO_2$.

In a detailed review and study of nanoparticles on $TiO_2$[18] it is assumed that the amount of charge transferred is significant due to the very high static dielectric constant of $TiO_2$ and that there will be a strong dipole at the metal-$TiO_2$ interface. It is also assumed that while the affected volume of charge depletion in $TiO_2$ will be substantially larger (on the order of $10^6$), the charge in the metal will remain at the interface due to the effective screening of metal electrons. For small metal particles (nano-range), the electric field will be dependent upon the metal-particle diameter. Nevertheless, the charge at the free metal surface should be negligible. Again, the unique interface between the metal particles and reduced rutile $TiO_2$ (particles grown from $TiO_2$) seems to indicate the opposite possibility in that the charge on the metal particles is not negligible.

Additionally, the charge is also of sufficient durability to provide cathodic protection from surface oxidation of the metal particles over an extended period of time. While no rhodium was detected after 6 months in the material (MS6) where the negative shift was observed at 4 ½ months, neither was any titanium detected. The carbon data suggested a carbon-based surface contamination on the sample, perhaps plasticizer from the plastic storage vial. However, the positive shift (after four months) in the Spichiger-Ulmann study is assumed due to a loss of the negative charge on the metal particles and could help to explain the contamination found on MS6. They also successfully re-instated the negative B.E. shift in the material (after it has shifted positive after four months) by subjecting it to another high temperature reduction cycle. This implies an association with the loss of conductivity of the $TiO_2$ with the loss of negative charge on the particle and provides further insight regarding the underlying mechanism.

It is therefore proposed that the unusual electron transfer is based on a unique metal-semiconductor interface that results from metal particles grown from the semiconductor itself. The interface significantly affects the Schottky barrier height and CNL resulting in a one-way open gateway for electron transfer (hence the term open gate phenomenon) from the reduced $TiO_2$ to the metal particles based on their difference in work function. The result is a significant and durable negative charge on the metal particles that infers cathodic protection from surface oxidation. In essence, the metal particles behave as negatively charged electrodes.

Any conjecture beyond this basic understanding is difficult. Many questions remain unanswered, particularly in regard to the material composition at the interface that enables the electron transfer to the metal particles. Nevertheless, the data is consistent indicating an unusual and previously unreported electron transfer phenomenon.

## V. CONCLUSION

The open gate phenomenon is based on readily observable and reproducible experimental results. Additional research is required to investigate the full nature of the interface and mechanism that results in the unusual negative charge on Group VIII metal particles grown from reduced rutile $TiO_2$. Once this is more fully understood, the open gate phenomenon may be applicable to other appropriate metal-reducible oxides and mixed metal oxides. Based on the preliminary findings, the open gate phenomenon has the potential to become the foundation for new technologies with energy applications including sensors, catalysts, and electrocatalysts, particularly for fuel cells.

## VI. ACKNOWLEDGEMENT

This research was performed at the former corporate R&D center of Diamond Shamrock Corporation and surface analysis labs of Case Western Reserve University and Ohio State University. The author gratefully acknowledges Dan Kalynchuk and Arnie Gordon, fellow researchers of the original investigations upon which this paper is based.